\documentclass[10pt]{article}

 \usepackage{amsmath,amsthm,amssymb}
 \usepackage{makeidx,epsfig,lscape}
 \usepackage{color,colortbl}
 \usepackage{fancyhdr}
 \usepackage{xcolor,pict2e}

 \renewcommand{\headrulewidth}{0pt}
 \renewcommand{\footrulewidth}{0.5pt}

 \definecolor{myaqua}{rgb}{0.0,0.5,0.55}
 \definecolor{lightaqua}{rgb}{0.75,0.95,0.95}

 \usepackage[colorlinks = true,
            linkcolor = myaqua,
            urlcolor  = blue,
            citecolor = myaqua]{hyperref}

\usepackage{caption}
\usepackage{floatrow}
\def\lin#1#2{\textcolor[rgb]{0.6,0.6,0.6}{\vspace*{#1mm} \hrule
   height 3 pt \vspace*{#2mm}}}
\def\bt{\begin{tabular}}
\def\et{\end{tabular}}
\def\and{\mbox{ and }}

\def\1{{\bf 1}}

 \def\sectionn#1{\refstepcounter{section}{\color{myaqua}

 \vskip 6mm

 \noindent\Large\bf\thesection. #1}

 \vskip 3mm}

 \def\boxx#1#2#3#4#5{
 {\linethickness{#4pt}\put(#1,#5){\color{myaqua}{\line(1,0){#3}}}}
 \multiput(#1,#2)(0,#4){2}{\line(1,0){#3}}
 \multiput(#1,#2)(#3,0){2}{\line(0,1){#4}}
  }

\usepackage{color}

\begin{document}

 \fancyhead[L]{\hspace*{-13mm}
 \bt{l}{\bf International Journal of Astronomy \& Astrophysics, 2019, *,**}\\
 Published Online **** 2019 in SciRes.
 \href{http://www.scirp.org/journal/*****}{\color{blue}{\underline{\smash{http://www.scirp.org/journal/****}}}} \\
 \href{http://dx.doi.org/10.4236/****.2014.*****}{\color{blue}{\underline{\smash{http://dx.doi.org/10.4236/****.2014.*****}}}} \\
 \et}
 \fancyhead[R]{\includegraphics{pic1.ps}}

 $\mbox{ }$

 \vskip 12mm

{ 

{\noindent{\huge\bf\color{myaqua}
Optical Spectroscopic Monitoring Observations of a T Tauri Star V409 Tau}}
%
\\[6mm]
{\large\bf Hinako Akimoto$^1$ and Yoichi Itoh$^1$}}
\\[2mm]
{ 
 $^1$ Nishi-Harima Astronomical Observatory, Center for Astronomy, 
University of Hyogo, 407-2 Nishigaichi, Sayo, Sayo, Hyogo 679-5313, Japan\\
Email: \href{mailto:akimoto@nhao.jp}{\color{blue}{\underline{\smash{akimoto@nhao.jp}}}}\\[1mm]

Received **** 2019
 \\[4mm]
Copyright \copyright \ 2019 by author(s) and Scientific Research Publishing Inc. \\
This work is licensed under the Creative Commons Attribution International License (CC BY). \\
\href{http://creativecommons.org/licenses/by/4.0/}{\color{blue}{\underline{\smash{http://creativecommons.org/licenses/by/4.0/}}}}\\
 \includegraphics{pic2.ps}

\lin{5}{7}

 { 
 {\noindent{\large\bf\color{myaqua} Abstract}{\bf \\[3mm]
 \textup{
We report the results of optical spectroscopic monitoring observations of a T Tauri star, V409 Tau. A previous photometric study indicated that this star experienced dimming events due to the obscuration of light from the central star with a distorted circumstellar disk. We conducted medium-resolution (R $\sim$ 10000) spectroscopic observations with 2-m Nayuta telescope at Nishi-Harima Astronomical Observatory. Spectra were obtained in 18 nights between November 2015 and March 2016. Several absorption lines such as Ca I and Li, and the H$\alpha$ emission line were confirmed in the spectra. The $I_{c}$-band magnitudes of V409 Tau changed by approximately 1 magnitude during the observation epoch. The equivalent widths of the five absorption lines are roughly constant despite changes in the $I_{c}$-band magnitudes. We conclude that the light variation of the star is caused by the obscuration of light from the central star with a distorted circumstellar disk, based on the relationship between the equivalent widths of the absorption lines and the  $I_{c}$-band magnitudes. The blue component of the H$\alpha$ emission line was dominant during the observation epoch, and an inverse P Cygni profile was observed in eight of the spectra. The time-variable inverse P Cygni profile of the H$\alpha$ emission line indicates unsteady mass accretion from the circumstellar disk to the central star. 
 }}}
 \\[4mm]
 {\noindent{\large\bf\color{myaqua} Keywords}{\bf \\[3mm]
Star formation; Pre-main sequence stars; T Tauri stars
}

 \fancyfoot[L]{{\noindent{\color{myaqua}{\bf How to cite this
 paper:}} Akimoto et al. (2019)
Optical Spectroscopic Monitoring Observations of a T Tauri Star V409 Tau.
 International Journal of Astronomy \& Astrophysics,*,***-***}}

\lin{3}{1}

\sectionn{Introduction}

{ \fontfamily{times}\selectfont
 \noindent 
Young stellar objects (YSOs) are generally variable stars. Some objects show periodic variability, while others exhibit irregular variability. Some objects also exhibit episodic variations. For example, FU Orionis outbursts represent the most extreme case of variability in these objects \cite{Bertout00}. Most members of this class have undergone outbursts of 4--6 magnitude in optical brightness \cite{Herbig77}; \cite{Hartmann96}. In \cite{Herbst94}, the authors proposed the movement of cool spots on the stellar surface due to stellar rotation, unsteady accretion of circumstellar material onto the star, and obscuration of light from the photosphere by a distorted circumstellar disk as causes of the light variation of YSOs.\\
~~~Thus far, numerous photometric studies on the variability of YSOs have been conducted. In \cite{Bouvier93}, the photometric monitoring of 24 YSOs was reported, and the authors observed one or two (quasi-) sinusoidal curve(s) with a period of 1.2 to 12 days in the optical light curves of 20 objects. They suggested spot movement on the stellar surface due to stellar rotation as the cause of the light variation. Also, numerous studies have also been reported on the variability of YSOs observed by spectroscopy. In \cite{Strassmeier94}, spectroscopic monitoring observations of a weak-lined T Tauri star, V410 Tau, were reported. The authors obtained spectra with a wavelength coverage of 3850 $\AA$ --9050 $\AA$ and a resolution of $\sim$12000 and identified hot spots on the photosphere by producing Doppler images of the object.\\
~~~Spectroscopic observations are useful for further understanding of the brightness variations of YSOs. \cite{Mekkaden07} conducted the photometric observation of HD 288313. This object showed light variations, which was considered to be attributed to change of coverage of cool spots. Distribution of the chromospheric active regions was examined by spectroscopic observations. The H$\alpha$ equivalent width shows rotational modulation only at occasional epochs. It was proposed that the chromospheric active regions spread across the stellar surface.\\
~~~AA Tau is another example of YSOs exhibiting light variation. In \cite{Bouvier07}, the authors reported high-resolution spectroscopic observations of AA Tau, which exhibits light variation due to obscuration of the photosphere by the inner edge of a magnetically-warped disk. When the star is faint, a red-shifted absorption line appears on the broad emission line of the Balmer series. The redshifted absorption line indicates infalling motion of cold material in front of the photosphere, suggesting an accretion flow of circumstellar material onto the photosphere. \\
~~~A classical T Tauri star, RW Aur A, is an irregular variable with a large amplitude. \cite{Petrov01} searched for periodicities in the variations of the brightness and colors of RW Aur A over three decades. With the spectroscopic observations they insisted the accretion of the magnetosphere. Study of the circumstellar environments of YSOs helps in the discussion of the formation of planets \cite{Rodriguez15}. In particular, objects that display large photometric dimming events caused by circumstellar disks give us the opportunity to study the evolution of the circumstellar environments of young stars.\\
~~~Photometric observations of V409 Tau were carried out with the Kilodegree Extremely Little Telescope North (KELT-North)\cite{Rodriguez15}. V409 Tau is a Class I\hspace{-.1em}I object \cite{Rebull10} with the spectral type of K8--M0 \cite{Rebull10} or M1.5 \cite{Luhman09}. They observed two separate dimming events; one from January 2009 to October 2010 and another from March 2012 until at least September 2013. In the latter event, the depth of the dimming was 1.4 magnitude in the $V$-band. They estimated the upper limit of the duration of the 2009 dimming event to 630 days. The interval between the beginnings of the two dimming events was 1130 days. By combining data from the Combined Array for Research in Millimeter Astronomy, the All Sky Automated Survey, the Catalina Real-time Transient Survey, and the Wide Angle Search for Planets, the authors constructed the spectral energy distribution, which was fitted with the star and disk SED model with a disk inclination angle of 81 degrees. They indicated that an almost edge-on circumstellar disk was distorted and that the dimming was caused by the obscuration of light from the central star by the distorted disk. \\
~~~V409 is a UX Orionis candidate \cite{Rodriguez15}. UX Ori type variability is observed in Herbig Ae/Be stars and some T Tauri stars with early K spectral types \cite{Bertout00}. The variability is characterized by large-amplitude light variations with no evident veiling or effective temperature variations. This is interpreted to be caused by variable obscuration by circumstellar dust. \cite{Bertout00} constructed an accretion disk model in which a large amount of accretion material heats gas in the circumstellar disk, making the disk flaring. They calculated the probability of observations of the star through the circumstellar disk. The result shows that high mass-accretion rate stars are more likely to be observed through the circumstellar disk than low mass-accretion rate stars. They suggested that accretion is driving UX Ori variability.\\
~~~In this paper, we present the spectroscopic monitoring observations of V409 Tau, which exhibits irregular variability. A previous photometric study \cite{Rodriguez15} suggested obscuration of the photosphere by a distorted circumstellar disk as the cause of the variability.

}

\renewcommand{\headrulewidth}{0.5pt}
\renewcommand{\footrulewidth}{0pt}

 \pagestyle{fancy}
 \fancyfoot{}
 \fancyhead{} 
 \fancyhf{}
 \fancyhead[RO]{\leavevmode \put(-90,0){\color{myaqua}H. Akimoto, et al.} \boxx{15}{-10}{10}{50}{15} }
 \fancyhead[LE]{\leavevmode \put(0,0){\color{myaqua}H. Akimoto, et al.}  \boxx{-45}{-10}{10}{50}{15} }
 \fancyfoot[C]{\leavevmode
 \put(0,0){\color{lightaqua}\circle*{34}}
 \put(0,0){\color{myaqua}\circle{34}}
 \put(-2.5,-3){\color{myaqua}\thepage}}

 \renewcommand{\headrule}{\hbox to\headwidth{\color{myaqua}\leaders\hrule height \headrulewidth\hfill}}

\sectionn{Observations and Data Analysis}

{ \fontfamily{times}\selectfont
\noindent
We conducted a series of spectroscopic observations of V409 Tau over 18 nights between November 2015 and March 2016. Observations were carried out with the medium- and low-dispersion long-slit spectrograph (MALLS) mounted on the nasmyth platform of the Nayuta telescope at Nishi-Harima Astronomical Observatory, Japan. With a grating of 1800 lines /mm and a 0.8$"$ slit, we obtained spectra with a wavelength resolution of $\sim$ 10000 between 6280 $\AA$ and 6720 $\AA$. This wavelength range was chosen for investigating the H$\alpha$ emission line which is an index of accretion, and metallic absorption lines originating from the photosphere. The exposure time ranged from 900 s --1200 s, and 1--17 spectra were taken each night. The goal was to reach S/N of 20 on good observation conditions.} Flat frames and comparison frames were acquired with a halogen lamp and an Fe-Ne-Ar lamp in the instrument, respectively. Dark frames were also taken. Details of the observations are shown in Table \ref{tab:log}.\\
\newlength{\myheightc}
\setlength{\myheightc}{0.4cm}
\begin{table}
  \caption{Observation date, exposure time, and S/N. }{%
  \begin{tabular}{cccccc}
      \hline
      \rule{0cm}{\myheightc}
      Date (JST) & Exp. Time (s) & S/N &Date & Exp. Time (s) & S/N\\ 
      \hline
      \rule{0cm}{\myheightc}
      2015 Nov 29 & 3600 & 23 & 2016 Jan 7	&6000	&11\\
      \rule{0cm}{\myheightc}
      2015 Nov 30 & 3600 & 19 & 2016 Jan 10	&8400	&16\\
      \rule{0cm}{\myheightc}
      2015 Dec 1   & 3600 & 25 &2016 Jan 12	&4800	&2\\
      \rule{0cm}{\myheightc}
      2015 Dec 4 & 3600 & 20 & 2016 Jan 13	&4800	&4\\
      \rule{0cm}{\myheightc}
      2015 Dec 7 &3600 & 29 &2016 Jan 16	&7200	&5 \\
      \rule{0cm}{\myheightc}
      2015 Dec 19	& 3600 & 8& 2016 Mar 7	&1200	&8\\
      \rule{0cm}{\myheightc}
      2015 Dec 27	& 6000 & 5 & 2016 Mar 11	&1200	&2\\
      \rule{0cm}{\myheightc}
      2015 Dec 29	&3600	&10& 2016 Mar 12	&4500	&11\\
      \rule{0cm}{\myheightc}
      2016 Jan 1	&20400	&39 &2016 Mar 26	&1800	&20\\
      \hline
    \end{tabular}}\label{tab:log}

\end{table}
~The image analysis software IRAF (Imaging Reduction and Analysis Facility) was used for image processing. First, the overscan and dark current were subtracted from the raw data. Next, we performed flat-fielding with the normalized flat frames, wavelength calibration and distortion correction with comparison frames, and background subtraction. The spectrum was extracted and binned based on the slit width. We combined the spectra acquired for each night, and finally, we normalized the continuum level of the spectrum. We shifted the wavelengths of the spectra so that the wavelengths of the five absorption lines described below matched to the wavelengths of the lines in vacuum.

%
\begin{figure}
 \begin{center}
  \includegraphics[width=14cm]{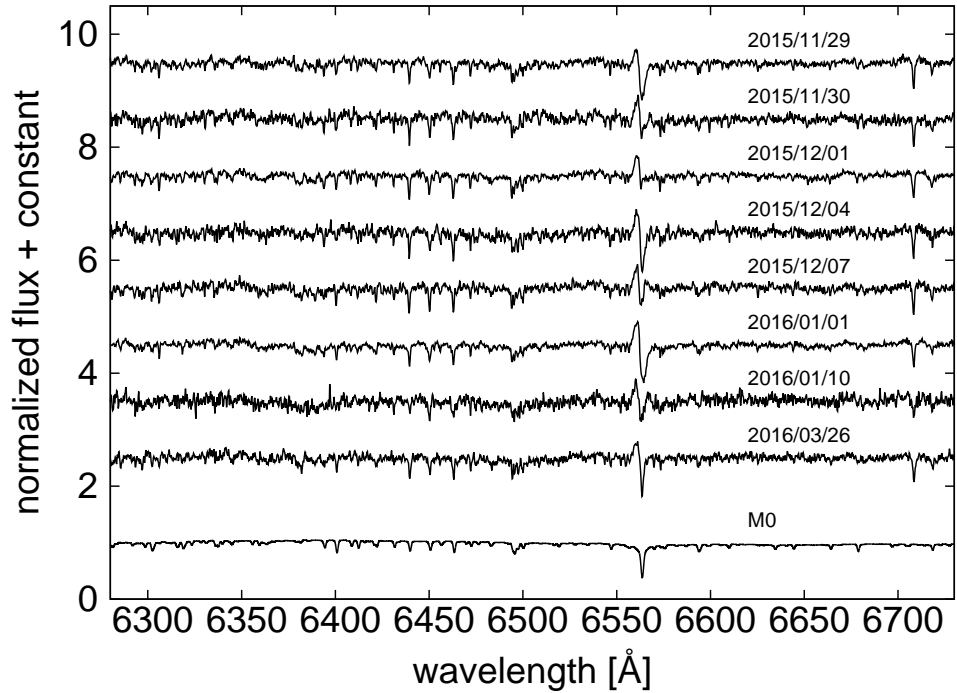} 
 \end{center}

\vspace*{+2.0cm}

 \caption{Optical spectra of V409 Tau. The horizontal axis represents the wavelength and the vertical axis presents the normalized flux + constant. Spectra with S/N ratios greater than 15 taken in eight nights are shown. Prominent absorption lines are Ca I at 6439 $\AA$, Ca I and Co I at 6450 $\AA$, Ca I at 6463 $\AA$, Li and Fe I at 6708 $\AA$, and Ca I at 6718 $\AA$. The H$\alpha$ emission line shows an inverse P Cygni profile that changes with time. A spectrum of an M0 dwarf (HD156274) is also shown  \cite{Bagnulo03}.}\label{fig:spec}
\end{figure}

}

\sectionn{Results}
{ \fontfamily{times}\selectfont
\noindent
We obtained spectra for 18 nights. The spectra exhibited the H$\alpha$ emission line and several absorption lines. The H$\alpha$ line displayed an inverse P Cygni profile. Among the spectra, those acquired for eight of the nights exhibited a signal-to-noise (S/N) ratio greater than 15, enabling clear confirmation of the absorption lines (Fig. \ref{fig:spec}). The equivalent widths of five absorption lines (Ca I at 6439 $\AA$, Ca I and Co I at 6450 $\AA$, Ca I at 6463 $\AA$, Li and Fe I at 6708 $\AA$, and Ca I at 6718 $\AA$) were measured using the SPLOT task in IRAF. The lines were fitted with a Gaussian function, and the error in the equivalent width was estimated from the root-mean-square of the continuum region adjacent to the line in the spectrum.\\
~~~Table \ref{tab:obs} shows the equivalent widths of the absorption lines and the $I_{c}$-band magnitudes of the object. The $I_{c}$-band magnitudes were taken from the database of the Kamogata /Kiso /Kyoto wide-field survey (KWS) \cite{Maehara14}. The light curve of V409 Tau is shown in the Figure \ref{fig:kws}. If the $I_{c}$-band magnitude was not available for a particular observation date, we used photometric data taken within one day of that date. We do not further discuss the spectrum taken in March 2016 because no photometric data are available within 30 days from the observation date. During the observation period, V409 Tau changed by 1 magnitude in the $I_{c}$-band; however, no significant changes were observed in the equivalent widths of the absorption lines.
\newlength{\myheighta}
\setlength{\myheighta}{0.4cm}

\begin{table}
  \caption{Observation date, $I_{c}$-band magnitude, and equivalent widths of the absorption lines. }{%
  \begin{tabular}{ccccccc}
      \hline
      \rule{0cm}{\myheighta}
      $Date$ & $I_{c} (mag)$ & $EW_{6439}(\AA)$ & $EW_{6450}(\AA)$ & $EW_{6463}(\AA)$ &  $EW_{6708}(\AA)$& $EW_{6718}(\AA)$  \\ 
      \hline
      \rule{0cm}{\myheighta}
      2015 Nov 29 & $11.45 \pm0.48$ & $0.38^{+0.07}_{-0.06}$ & $0.39^{+0.13}_{-0.09}$ & $0.59^{+0.19}_{-0.19}$ & $0.53^{+0.10}_{-0.06}$ & $0.29^{+0.07}_{-0.12}$  \\
      \rule{0cm}{\myheighta}
      2015 Nov 30 & $11.19 \pm0.15$ & $0.42^{+0.11}_{-0.09}$ & $0.37^{+0.11}_{-0.08}$ & $0.49^{+0.16}_{-0.13}$ & $0.56^{+0.14}_{-0.11}$ & $0.26^{+0.79}_{-0.10}$   \\
      \rule{0cm}{\myheighta}
     2015 Dec 1   &$11.19 \pm0.15$ & $0.43^{+0.05}_{-0.05}$ & $0.41^{+0.08}_{-0.07}$ & $0.53^{+0.08}_{-0.08}$ & $0.59^{+0.07}_{-0.06}$ & $0.68^{+0.20}_{-0.22}$ \\
      \rule{0cm}{\myheighta}
     2015 Dec 4	&$10.97 \pm0.12$ & $0.51^{+0.19}_{-0.11}$ & $0.57^{+0.33}_{-0.21}$ & $0.52^{+0.12}_{-0.15}$ & $0.59^{+0.09}_{-0.08}$ & $0.27^{+0.71}_{-0.02}$  \\
      \rule{0cm}{\myheighta}
      2015 Dec 7	&$11.14 \pm0.13$ & $0.41^{+0.10}_{-0.08}$ & $0.41^{+0.10}_{-0.09}$ & $0.48^{+0.40}_{-0.11}$ & $0.63^{+0.09}_{-0.08}$ & $0.36^{+0.50}_{-0.04}$ \\
      \rule{0cm}{\myheighta}
     2016 Jan 1	& $11.59\pm0.24$ & $0.38^{+0.04}_{-0.04}$ & $0.45^{+0.01}_{-0.10}$ & $0.45^{+0.07}_{-0.07}$ & $0.43^{+0.04}_{-0.04}$ & $0.18^{+0.03}_{-0.08}$ \\
      \rule{0cm}{\myheighta}
    2016 Jan 10	& $12.01 \pm0.36$ & $0.16^{+0.26}_{-0.05}$ & $0.36^{+0.11}_{-0.09}$ & $0.54^{+0.20}_{-0.23}$ & $0.50^{+0.27}_{-0.27}$ & $0.35^{+0.04}_{-0.27}$  \\
      \rule{0cm}{\myheighta}
    2016 Mar 26& - & $0.45^{+0.11}_{-0.09}$ & $0.41^{+0.25}_{-0.13}$ & $0.43^{+0.36}_{-0.10}$ & $0.59^{+0.13}_{-0.11}$ & $0.27^{+0.46}_{-0.13}$\\
      \hline
    \end{tabular}}\label{tab:obs}

\end{table}
\begin{figure}
 \begin{center}
  \includegraphics[width=11cm]{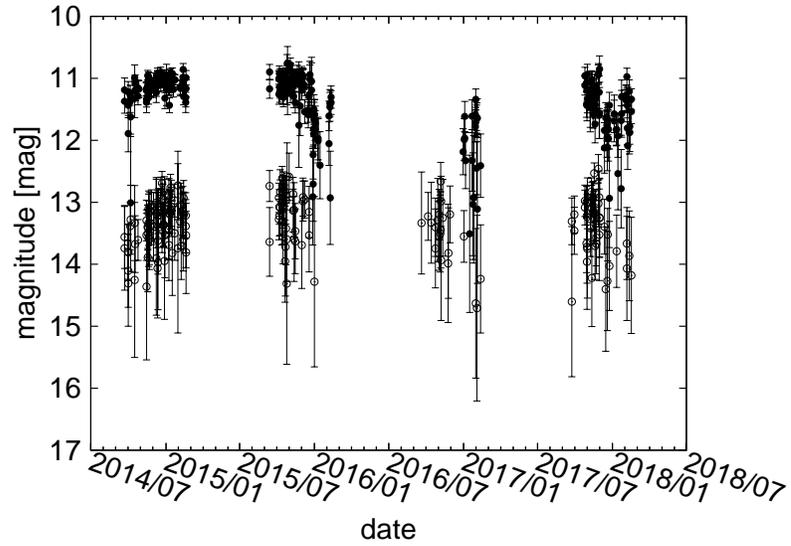} 
 \end{center}

\vspace*{+2.0cm}

 \caption{KWS light curve of V409 Tau. The filled circles indicate the $I_{c}$-band magnitudes. The open circles indicate the $V$-band magnitudes.}\label{fig:kws}
\end{figure}

}

\sectionn{Discussion}
{ \fontfamily{times}\selectfont
\noindent
\subsection{Absorption lines}
V409 Tau showed a 1-magnitude variation in the $I_{c}$-band during the observation epoch. To specify the cause of the light variation, we constructed three simple models, in which the variation of the equivalent widths of the absorption lines was investigated. Since the SED of V409 Tau indicates little contribution of the circumstellar disk in the optical wavelength range \cite{Rodriguez15}, we assume that the absorption lines in the optical region originate from the photosphere.
\\
~~~First, we investigated the possibility that the luminosity changes resulted from the coverage of cool spots on the stellar surface. The spots have a temperature approximately 1000 K lower than the temperature of the photosphere \cite{Bouvier93}. Assuming a constant stellar radius and an effective temperature of 3800 K \cite{Pecaut13}, the radiation of the central star, $I_{\lambda}$, is the sum of the blackbody radiation, $B_{\lambda}(T)$, of the normal photosphere and that of the cool spots, as follows;
\begin{equation}
I_{\lambda} = B_{\lambda}(3800) \cdot ( 1 - \beta ) + B_{\lambda}(2800) \cdot \beta,
\end{equation}
where $\beta$ (0 $\le$ $\beta$ $\le$ 1) represents the coverage of the cool spots. The calculated radiation was multiplied by the transmittance of the $I_{c}$-band filter \cite{Bessell90}. We set $\beta$ = 0 for the brightest epoch (11 magnitude in the $I_{c}$-band) and calculated the $I_{c}$-band magnitude by changing $\beta$. For the faintest epoch (12 magnitude in the $I_{c}$-band), we set $\beta$ = 0.5. The equivalent width of an absorption line is the sum of the equivalent widths of the absorption lines for the 3800-K and 2800-K spectra.
\begin{equation}
EW = EW(3800) \cdot ( 1 - \beta ) + EW(2800) \cdot \beta
\end{equation}
The model spectra were calculated using the BT-NextGen model \cite{Allard09}. We assumed the surface gravity to be log g = 3.5 and the metallicity to be the same as the Sun. The spectra for temperatures of 3800-K and 2800-K were calculated and combined with the coverage factor, $\beta$, and we smoothed the model spectrum to the spectral resolution of the observed spectra. Table \ref{tab:cool} shows the calculated equivalent widths of the absorption lines. The equivalent widths decrease or are constant with increasing the spot coverage.
\newlength{\myheight}
\setlength{\myheight}{0.4cm}

\begin{table}
  \caption{Equivalent widths of the absorption lines expected for luminosity changes caused by cool spots on the stellar surface.}{%
  \begin{tabular}{ccccccc}
      \hline
      $\beta$ & 0.0 & 0.1 & 0.2 & 0.3 & 0.4 & 0.5  \\ 
      \hline
     \rule{0cm}{\myheight}
      $I_{c}$ (mag) & 11.00 & 11.35 & 11.45 & 11.56 & 11.83 & 11.99 \\
     \hline
     \rule{0cm}{\myheight}
      $EW_{6439}$ $(\AA)$	&1.04& 	0.99& 	0.94& 	0.89& 	0.84 &	0.79 \\
     \rule{0cm}{\myheight}
      $EW_{6450}$ $(\AA)$ &	0.25 &	0.23 &	0.21 &	0.19 &	0.17 &	0.15 \\
     \rule{0cm}{\myheight}
      $EW_{6463}$ $(\AA)$&	0.88 &	0.83 &	0.79 &	0.74 &	0.69& 	0.65 \\
     \rule{0cm}{\myheight}
      $EW_{6708}$ $(\AA)$&	0.79& 	0.79 &	0.79& 	0.79 &	0.79 &	0.78 \\
     \rule{0cm}{\myheight}
      $EW_{6718}$ $(\AA)$&	0.30 &	0.30 &	0.29 &	0.29 &	0.29 &	0.29 \\
     \hline
    \end{tabular}}\label{tab:cool}

\end{table}
We next considered unsteady material accretion from the circumstellar disk onto the photosphere. When the material accretion rate increases, the boundary layer connects to the inner part of the disk \cite{Popham93} and emits continuum light in the ultraviolet and optical wavelengths \cite{Bertout88}. We assumed constant intensities of the photosperic absorption lines. When the intensity of continuum increases by $\alpha$, the equivalent width $EW'$ is $S / (I + \alpha) = EW / (1 + V)$, where $I$ is the intensity of continuum, $S$ is the intensity of absorption line, and $V = \alpha / I$. We set $V = 0$ and used the 3800 K spectrum in the BT-NextGen model when the star is 12 magnitude. The $EWs$ of the absorption lines were calculated until $V$ increased and the $I_{c}$-band magnitude reached 11 magnitude.\\
~~~We also considered the obscuration of light from the photosphere by the distorted disk. AA Tau-like variables show periodic dips in their light curves, which are thought to be caused by the obscuration of light from the photosphere caused by warps in the inner disk. The periods of the dips are typically five to ten days, corresponding to a distance of 0.1 AU between the central star and the inner disk for Keplerian rotation \cite{Schneider18}. In the case of AA Tau, the luminosity drops by 1.2 magnitude with a period of 8.2 days \cite{Bouvier99}, \cite{Bouvier07}. It is proposed that the inner disk is distorted by the magnetic field and thus obscures the photosphere. As light from the central star passes through the disk, the intensity decreases uniformly within a certain wavelength range. As the absorption line weakens, the adjacent continuum light also weakens. As a result, the equivalent width of the absorption line does not change when the photosphere is obscured. \\}
\begin{figure}[htbp]
\begin{minipage}{7.3cm}
\begin{center}
\includegraphics[height=5.3cm]{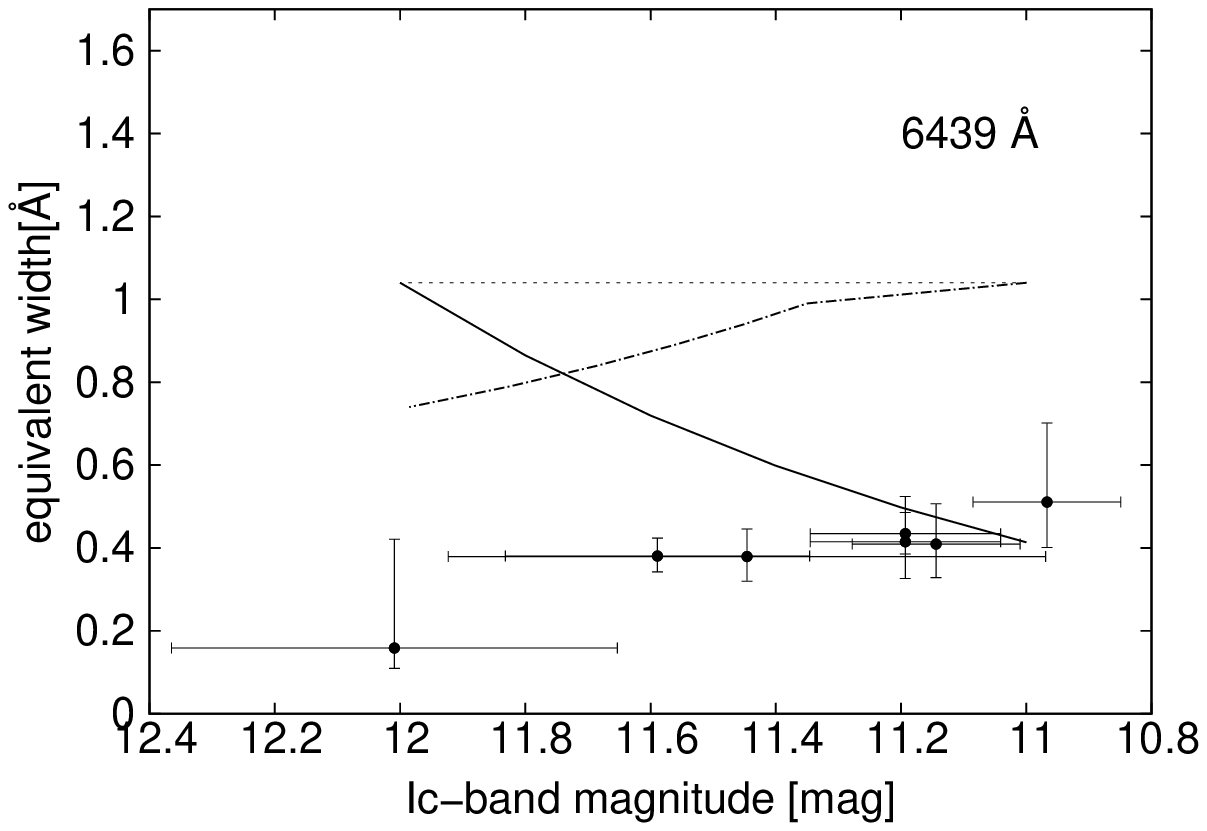}
\end{center}
\end{minipage}\quad
\begin{minipage}{7.3cm}
\begin{center}
\includegraphics[height=5.3cm]{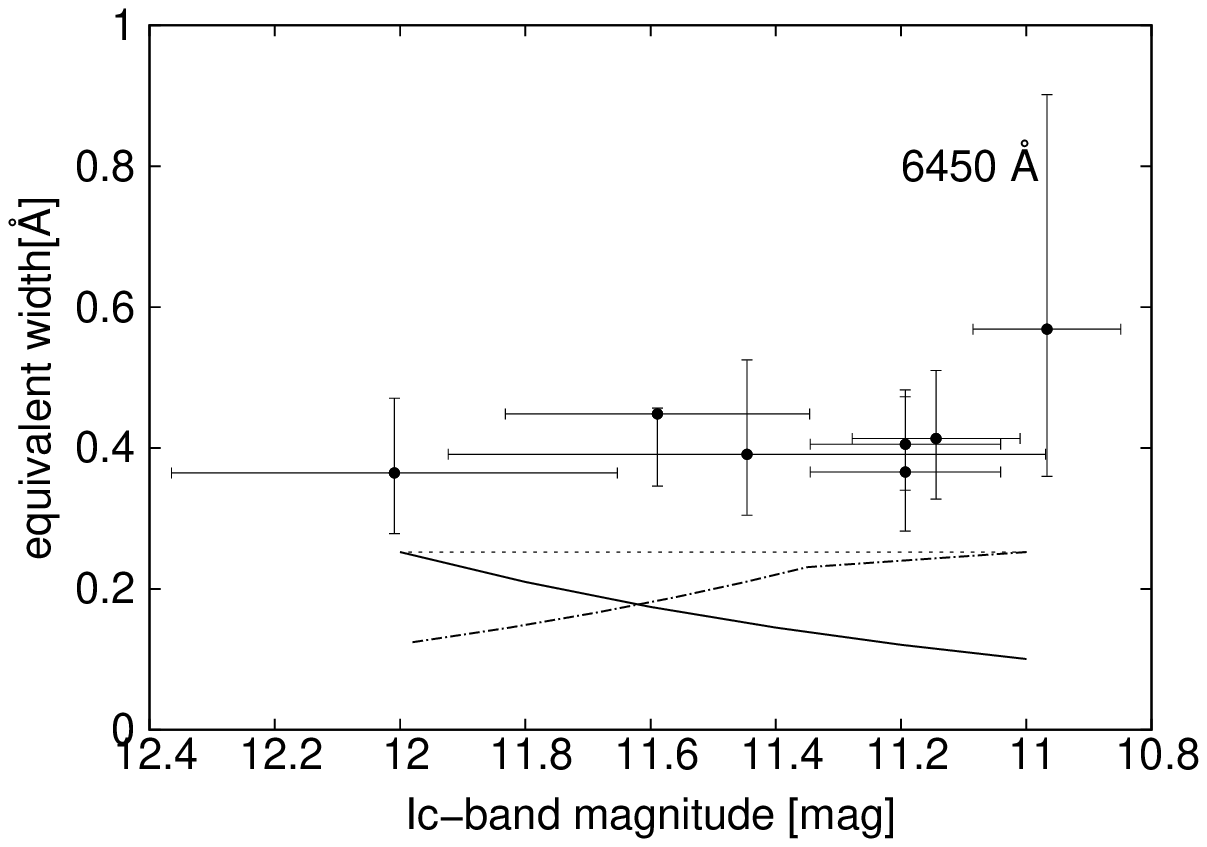}
\end{center}
\end{minipage}\quad

\begin{minipage}{7.3cm}
\begin{center}
\includegraphics[height=5.3cm]{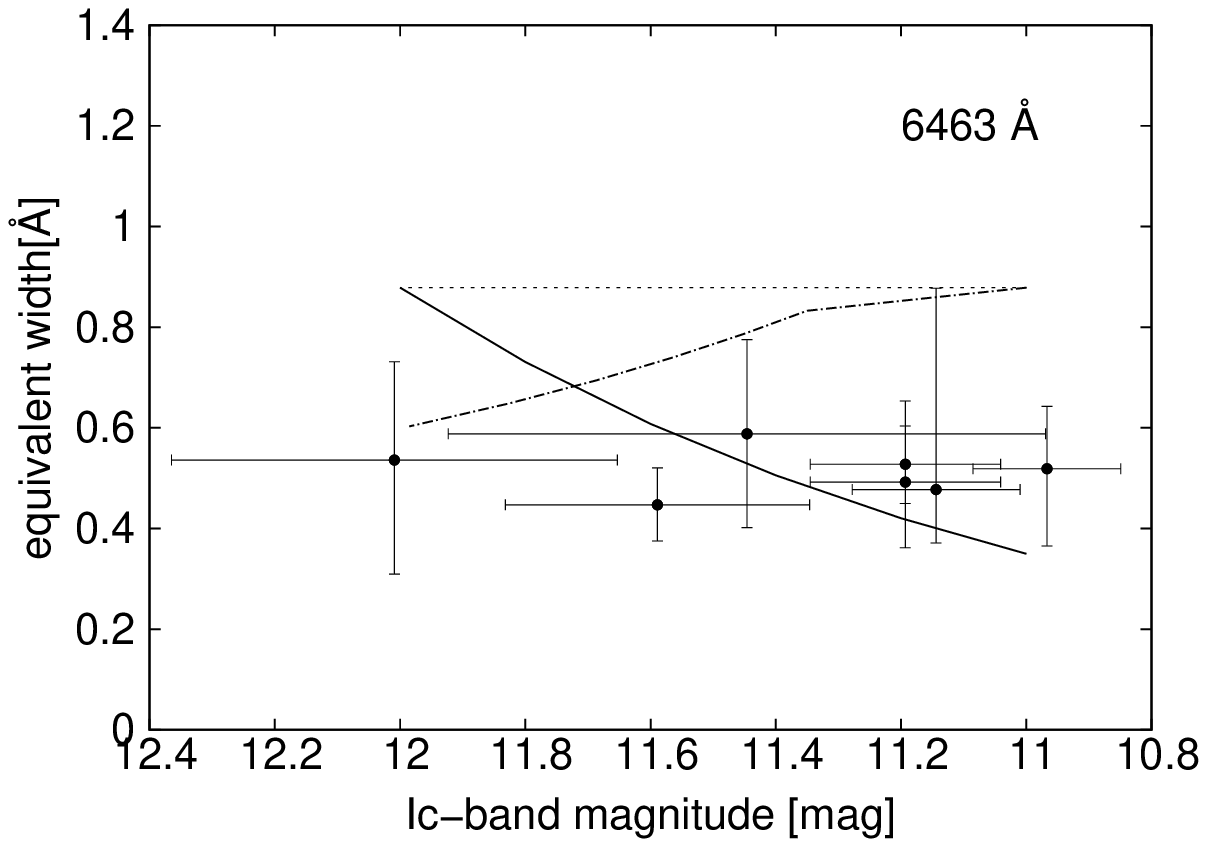}
\end{center}
\end{minipage}\quad
\begin{minipage}{7.3cm}
\begin{center}
\includegraphics[height=5.3cm]{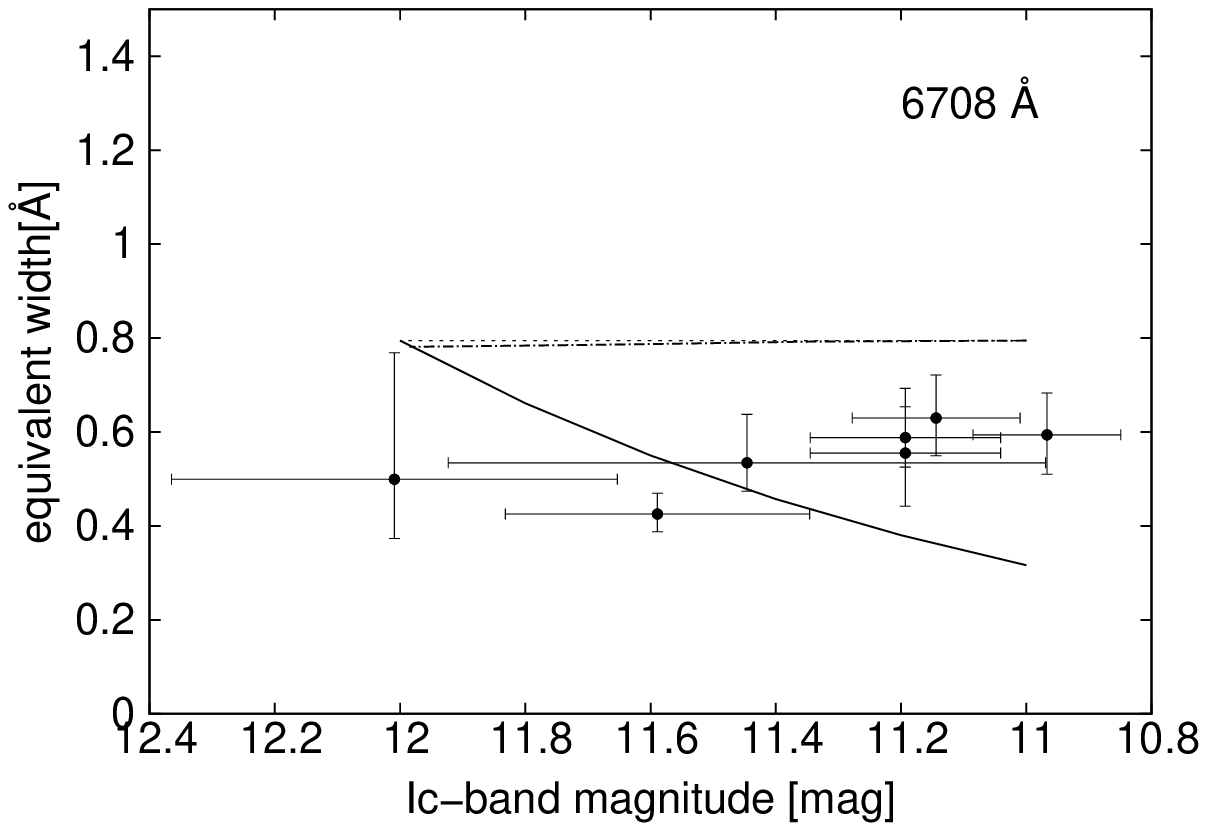}
\end{center}
\end{minipage}\quad

\begin{minipage}{7.3cm}
\begin{center}
\includegraphics[height=5.3cm]{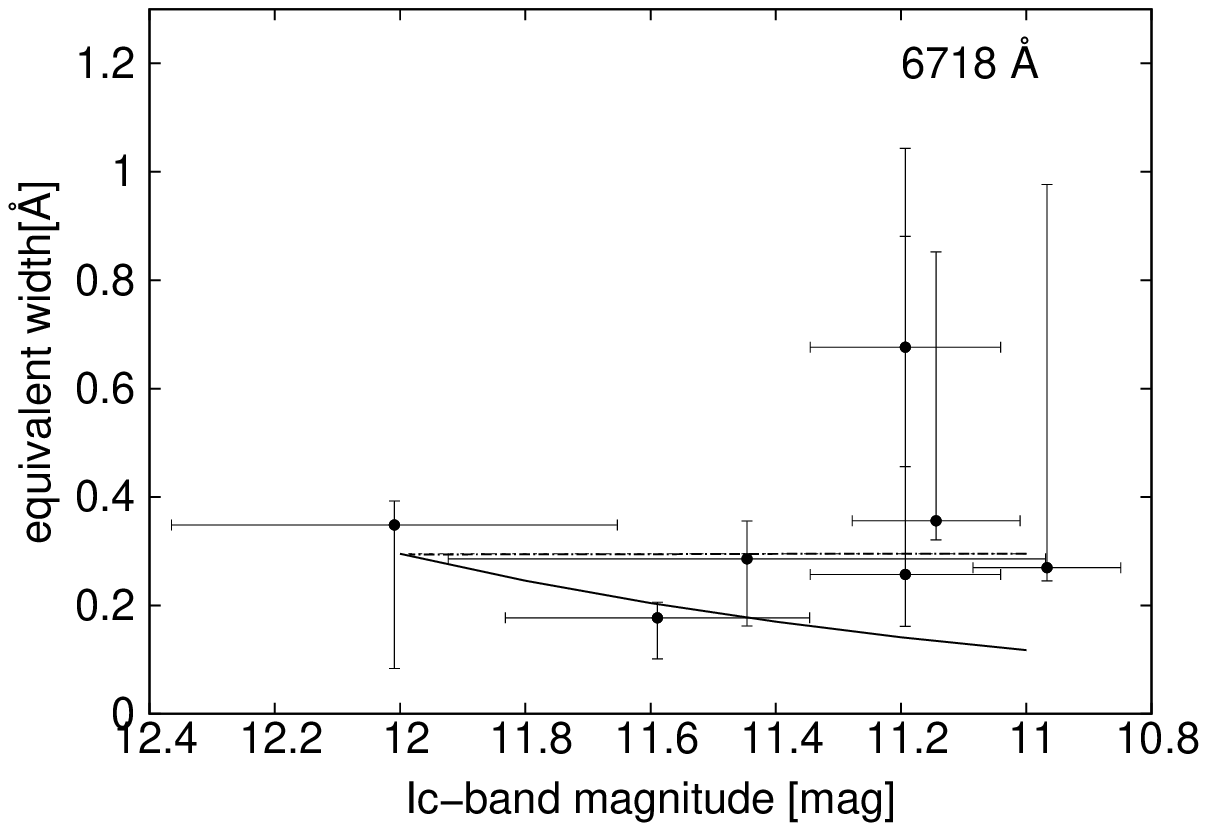}
\end{center}
\end{minipage}\quad
\vspace{1.3cm}


\caption{Relationship between the equivalent widths of the absorption lines and the  $I_{c}$-band magnitudes. The equivalent widths of the five absorption lines at 6439 $\AA$, 6450 $\AA$, 6463 $\AA$, 6708 $\AA$, and 6718 $\AA$ were measured (filled circles). Dashed-dotted line: equivalent widths expected if the luminosity changes result from variations in the coverage of cool spots on the stellar surface. Solid line: equivalent widths expected if the luminosity changes result from unsteady material accretion from the circumstellar disk onto the photosphere. Dotted line: equivalent widths expected if the luminosity changes result from  obscuration by the distorted disk. The equivalent widths of the obscuration model are shifted to match the equivalent width of the cool spot model at 11 magnitude. The measured equivalent widths are roughly constant for all absorption lines regardless of the  $I_{c}$-band magnitude.}\label{fig:abs}
\end{figure}
~~Figure \ref{fig:abs} shows the relationship between the equivalent widths of the absorption lines and the broadband magnitudes. The observed equivalent widths are roughly constant at all magnitude. In the case of the cool spots, the equivalent widths decrease slightly or are constant with decreasing luminosity. In the case of accretion, the equivalent widths increase with decreasing luminosity. In the case of the disk, the equivalent widths are constant. Figure \ref{fig:abs} indicates that the cause of the 2015 dimming event is the case where the coverage of cool spots on the stellar surface has changed, or the case where the light from the photosphere is obscured by a distorted disk. V409 Tau has a periodic photometric variation of 4.754 days, corresponding to the rotation period of the photosphere with star spots \cite{Xiao12}. This period is significantly shorter than the duration of the 2015 dimming event. The cool spot model indicates that the large portion of the photosphere is covered by the spot when the star was faint. The temperature of the spot is as low as 2800 K, so that the spectrum would show absorption features at 6650 $\AA$ and 6680 $\AA$. However, no clear absorption features appeared in the observed spectra. Therefore we claim the occultation by the disk as the cause of the light variation.\\
\begin{figure}
 \begin{center}
  \includegraphics[width=10cm]{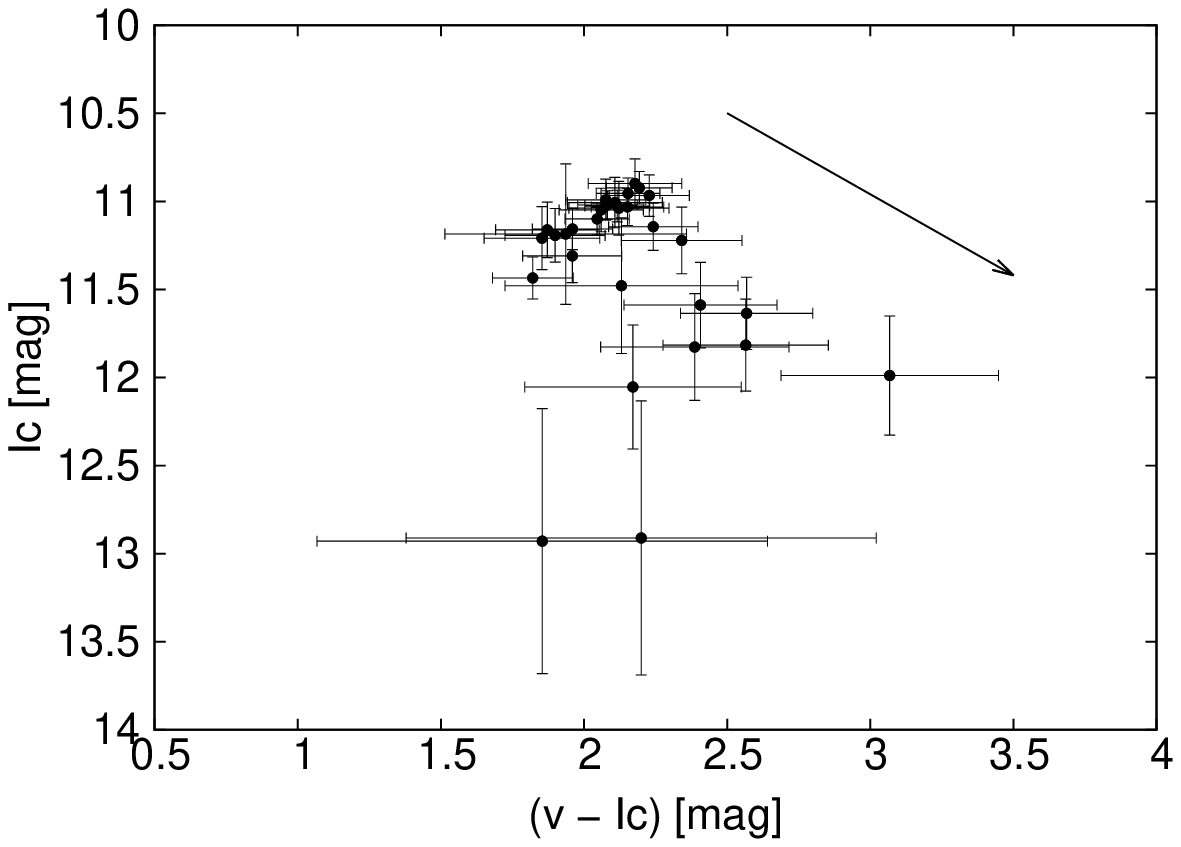} 
 \end{center}

\vspace*{+2.0cm}

 \caption{$(V-I_{c}, I_{c})$ color-magnitude diagram. We used $ I_ {c} $-band photometric data from KWS and $ V $ -band photometric data from ASAS-SN. The arrow  indicates interstellar extinction. Photometric variation can be interpreted as the intersteller extinction.}\label{fig:v-i}
\end{figure}
~~~Figure \ref{fig:v-i} is a $(V-I_{c}, I_{c})$ color-magnitude diagram. The $I_{c}$-band data are taken from the KWS database. We acquired $V$-band photometric data from ASAS-SN \cite{Shappee14}  taken on the same day at the same day as the KWS data. The arrow in the figure indicates interstellar extinction. Photometric variation can be interpreted as the intersteller extinction. This further supports that the dimming event occurred due to the obscuration by the disk.\\
~~~\cite{Rodriguez15} presented the light variation of V409 Tau.  They reported that the star experienced two dimming events in 2009--2010 and 2012--2013. The authors discussed the number of obscuring objects and their distances from the central star. In the discussion, they presented the possibility that obscuring objects of 2009 and 2012 are different. This is because V409 Tau did not experience significant dimming in 2002 and 2006. Other occultation happened in the 1960s \cite{Romano75}. They suggested that the obscuring object is located at farther 10.7 AU from the central star if the dimming events are attributed to different obscuring objects. They argued that if it is a single feature, it implies that the duration and depth is changing. In this case, the distance between the central star and the obscuring object was determined as 1.7 AU.\\
~~~In this paper, the duration of the dip cannot be precisely estimated due to the sparsity of the photometric data. According to the KWS data, the $I_{c}$-band magnitude of the star increased to 11.2 mag on September 22, 2014 and then varied within the range of 0.5 mag. The magnitude dropped to 12.9 mag on December 28, 2015 and then returned to 11.2 mag on October 26, 2017. We assume a single obscuration object, with dimming events ending on September 22, 2014 and October 26, 2017. If we adopt 0.57 $M_{\odot}$ and 1.11 $R_{\odot}$ \cite{Rodriguez15} for the mass and radius of the central star, respectively, the distance between the center star and the obscuration object is 1.76 AU. In our observations, the maximum duration of the dimming event was $\sim$668 days (December 28, 2015--October 26, 2017), similar to that of the 2009--2010 dimming event. However, the beginning of the 2012--2013 dimming event is not consistent with the orbital period of the obscuration object. For a precise determination of the shape and number of obscuring objects and the distance from the central star, more frequent photometric observations are required.\\
\subsection{H$\alpha$ emission line}
The H$\alpha$ emission line of V409 Tau shows an asymmetric profile with time variation. The H$\alpha$ line had an S/N \verb|>| 5 in the spectra acquired for 13 nights. Among these spectra, nine displayed an inverse P Cygni profile. (Table \ref{tab:pcyg})\\
\newlength{\myheightd}
\setlength{\myheightd}{0.4cm}

\begin{table}
  \caption{Observation date and equivalent widths of the blueward and redward components in the H$\alpha$ emission line.}{%
  \begin{tabular}{cccc}
      \hline
      \rule{0cm}{\myheightd}
date (JST)&	$EW_{\rm blue}$ $(\AA)$&$EW_{\rm red}$ $(\AA)$	&inverse P Cygni\\
      \hline
      \rule{0cm}{\myheightd}
2015 Nov 29	&$0.63^{+0.23}_{-0.16}$ & $0.01^{+0.10}_{-0.01}$ &  	yes\\
      \rule{0cm}{\myheightd}
2015 Nov 30	&$1.22^{+0.33}_{-0.25}$ & $0.32^{+0.28}_{-0.17}$ &  	yes\\
      \rule{0cm}{\myheightd}
2015 Dec 1	&$0.99^{+0.22}_{-0.16}$ & $0.06^{+0.09}_{-0.04}$ &  	yes\\
      \rule{0cm}{\myheightd}
2015 Dec 4	&$0.90^{+0.38}_{-0.25}$ & $0.18^{+0.20}_{-0.13}$ &  	yes\\
      \rule{0cm}{\myheightd}
2015 Dec 7	&$1.03^{+0.33}_{-0.25}$ & $0.43^{+0.38}_{-0.25}$ &  	yes\\
      \rule{0cm}{\myheightd}
2015 Dec 19	&$2.59^{+1.18}_{-0.85}$ & $0.88^{+1.02}_{-0.54}$ &  	no\\
      \rule{0cm}{\myheightd}
2015 Dec 29	&$1.58^{+0.97}_{-0.65}$ & $0.59^{+0.78}_{-0.39}$ &  	no\\
      \rule{0cm}{\myheightd}
2016 Jan 1	&$1.28^{+0.23}_{-0.17}$ & $0.03^{+0.09}_{-0.02}$ &  	yes\\
      \rule{0cm}{\myheightd}
2016 Jan 7	&$3.64^{+0.66}_{-0.52}$ & $0.42^{+0.54}_{-0.30}$ &  	no\\
      \rule{0cm}{\myheightd}
2016 Jan 10	&$1.07^{+0.55}_{-0.40}$ & $0.48^{+0.43}_{-0.27}$ &  	yes\\
      \rule{0cm}{\myheightd}
2016 Mar 7	&$1.15^{+0.71}_{-0.54}$ & $0.60^{+0.65}_{-0.43}$ &  	yes\\
      \rule{0cm}{\myheightd}
2016 Mar 12	&$1.42^{+0.54}_{-0.40}$ & $0.58^{+0.53}_{-0.36}$ &  	no\\
      \rule{0cm}{\myheightd}
2016 Mar 26	&$0.78^{+0.31}_{-0.24}$ & $0.12^{+0.22}_{-0.10}$ &  	yes\\
      \hline

    \end{tabular}}\label{tab:pcyg}

\end{table}
In \cite{Alcala93}, the authors reported on low- and high-resolution spectroscopy of a T Tauri star, T Cha. The H$\alpha$ emission line profile changed from pure emission to an inverse P Cygni profile in less than one day. The authors claimed that a red-shifted absorption component appeared in the optically thin H$\alpha$ emission when an episodic mass accretion was caused. \cite{Edwards94} claimed that the redshifted absorption arises from an accretion flow from the inner disk to the stellar surface along the magnetospheric field lines. As suggested by the relatively high projected rotational velocity of the star \cite{Franchini92}, T Cha is seen in an almost edge-on geometry, as is the case for V409 Tau \cite{Rodriguez15}. We propose that the redshifted absorption for V409 Tau arises from the accretion flow and is overlaid on the broad emission line.\\
\begin{figure}
 \begin{center}
  \includegraphics[width=10cm]{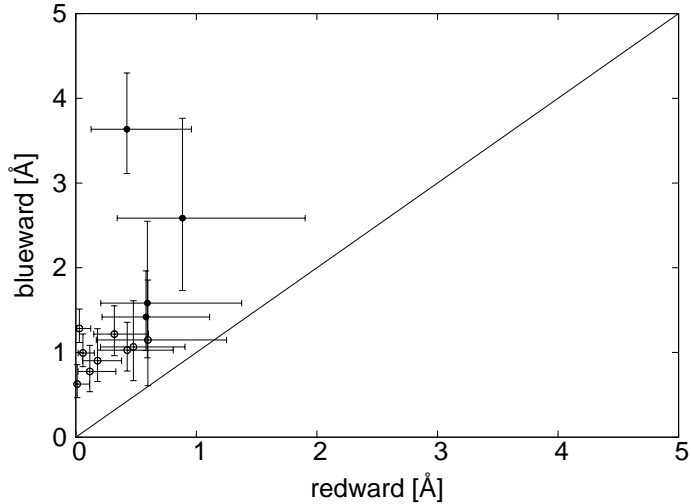} 
 \end{center}

\vspace*{+1.0cm}
\caption{Equivalent widths of the redward and blueward components of the H$\alpha$ emission line of V409 Tau. The open circle indicates the inverse P Cygni profile. The blueward components are dominant for all spectra, and the inverse P Cygni profile appeared when the H$\alpha$ emission was weak.}\label{fig:h}
\end{figure}
~~~In \cite{Reipurth96}, the shape of the H$\alpha$ emission line of 63 YSOs was analyzed. The authors measured the equivalent widths of the blueward and redward components of the line. For the relatively weak emission line stars [10 $\AA$ \verb|<| EW(H$\alpha$) \verb|<| 70 $\AA$], the equivalent widths of the blueward and redward components are comparable. In contrast, among the strong emission line stars [EW(H$\alpha$) $\geq$ 70 $\AA$], the redward components are stronger than the blueward components for many objects.\\
~~~We measured the equivalent widths of the redward and blueward components of the H$\alpha$ emission line of V409 Tau between -500 km/s and +500 km/s (Figure \ref{fig:h}). We set 6563$\AA$ to 0 km/s. We defined redward component as the velocity range from 0 km/s to +500 km/s and blueward component from 0 km/s to -500 km/s. We did not measure the equivalent width of the absorption line. The total equivalent width of the redward and blueward components of V409 Tau varies from 0.64 $\AA$ to 4.08 $\AA$. For the H$\alpha$ emission line of V409 Tau, the blueward component is dominant. The equivalent widths of both the blueward and redward components vary with time. The time-variable inverse P Cygni profile of the H$\alpha$ emission line of V409 Tau indicates unsteady mass accretion from the circumstellar disk to the central star. Based on the correlation between the mass of T Tauri stars and the mass accretion rate \cite{Gregory06}, the mass accretion rate of V409 Tau is estimated to be $10 ^ {- 8}$ $M_{\odot}$/yr. Assuming that the inner radius of the accretion disk is five times the stellar radius \cite{Gullbring98}, $R_{s}$ = 1.11 $R _{\odot}$, and $M_{s}$ = 0.57 $M_{\odot}$ \cite{D'Orazi11}; \cite{Rodriguez15}, we determined that the accretion luminosity is 0.127 $L_{\odot}$. Thus, the luminosity of the object increases by 0.4 magnitude when the accretion phenomenon occurs. Because continuum excesses caused by accretion shocks primarily contribute to the blue and ultraviolet flux \cite{Hartmann16}, we expect that the light variation due to unsteady mass accretion is less than 0.4 magnitude in the $I_{c}$-band.\\
\begin{figure}
 \begin{center}
  \includegraphics[width=10cm]{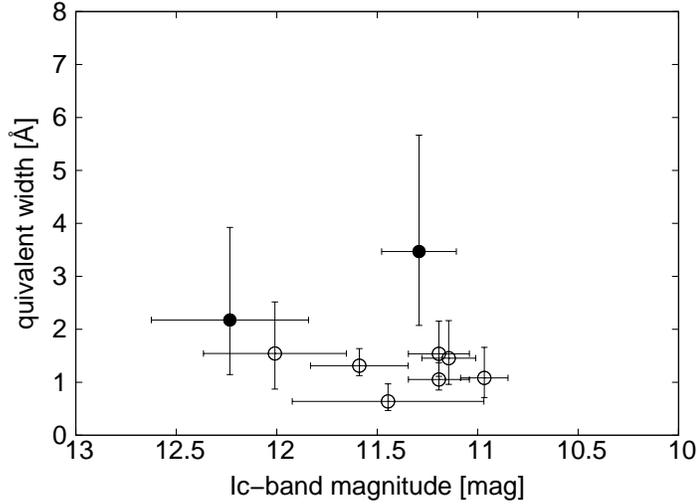} 
 \end{center}

\vspace*{+1.0cm}
\caption{Equivalent widths of the H$\alpha$ emission line of V409 Tau and $I_{c}$-band magnitudes. Most of the H$\alpha$ lines show the inverse P Cygni profile (open circles).}\label{fig:ew-i}
\end{figure}
~~~Figure \ref{fig:ew-i} shows the relationship between the $I_{c}$-band magnitudes and the equivalent widths of H$\alpha$ emission line. Except for one data, one may find a tendency that the equivalent widths increase with decreasing the broad-band brightness. This implies light variation due to veiling. However, most of the H$\alpha$ lines show the inverse P Cygni profile. Since we are measuring only the emission line part, it is necessary to construct a detailed model that takes self-absorption into account (e.g. \cite{Calvet92}).

\sectionn{Summary}

{ \fontfamily{times}\selectfont
\noindent
We conducted optical spectroscopic monitoring of a T Tauri star V409 Tau. Medium-resolution spectra were obtained over 18 nights in 2015 and 2016. While the $I_{c}$-band magnitude of the object changed by 1 magnitude during the observations, the equivalent widths of the absorption lines remained nearly constant. We constructed three simple models to investigate the variation of the equivalent widths of the absorption lines. We concluded that the light variation arises from the distorted disk. If only one obscuring feature is present, the distance between the central star and the obscuring feature is estimated as 1.76 AU. The H$\alpha$ emission line profile of V409 Tau showed a time-variable inverse P Cygni profile, indicating unsteady mass accretion from the circumstellar disk to the central star.\\

}

}

 {\color{myaqua}

 \vskip 6mm

 \noindent\Large\bf Acknowledgments}

 \vskip 3mm

{ \fontfamily{times}\selectfont
 \noindent
H. A. is supported by the Iue Memorial Foundation.  
 {\color{myaqua}

}}

\end{document}